\documentclass[twocolumn,prl,aps,showpacs,floatfix,superscriptaddress,preprintnumbers,amsmath,amssymb]{revtex4}

\usepackage{graphicx}

\begin{document}

\renewcommand{\floatpagefraction}{0.5}

\title{Morphology of the Superconducting Vortex Lattice in Ultra-Pure Niobium}

\author{S. M\"uhlbauer}

\affiliation{Technische Universit\"at M\"unchen, Physik Department E21, D-85748 Garching, Germany}

\affiliation{Forschungsneutronenquelle Heinz Maier-Leibnitz, D-85747 Garching, Germany}

\author{C. Pfleiderer}

\affiliation{Technische Universit\"at M\"unchen, Physik Department E21, D-85748 Garching, Germany}

\author{P. B\"oni}

\affiliation{Technische Universit\"at M\"unchen, Physik Department E21, D-85748 Garching, Germany}

\author{M. Laver}

\affiliation{NIST Center for Neutron Research, Gaithersburg, Maryland 20899 USA}

\affiliation{University of Maryland, College Park, Maryland, 20742 Maryland USA}

\author{ E. M. Forgan}

\affiliation{School of Physics and Astronomy, University of Birmingham, Birmingham, UK}

\author{D. Fort}

\affiliation{Metallurgy and Materials Science, University of Birmingham, Birmingham, UK}

\author{U. Keiderling}

\affiliation{Helmholtz Zentrum Berlin, BENSC, D-14109 Berlin, Germany}

\author{G. Behr}

\affiliation{Leibnitz-Institut f\"ur Festk\"orper- und Werkstoffforschung IFW, D-01069 Dresden, Germany}

\date{\today}

\begin{abstract}
The morphology of the superconducting flux line lattice (FLL) of Nb comprises gradual variations with various lock-in transitions and symmetry breaking rotations. We report a comprehensive small-angle neutron scattering study of the FLL in an ultra-pure single crystal of Nb as a function of the orientation of the applied magnetic field. We attribute the general morphology of the FLL and its orientation to three dominant mechanisms. First, non-local contributions, second, the transition between open and closed Fermi surface sheets and, third, the intermediate mixed state between the Meissner and the Shubnikov phase.
\end{abstract}


\vskip2pc

\maketitle

In the conventional Ginzburg-Landau theory of superconductivity the formation of a flux line lattice (FLL) -- the so-called Shubnikov phase -- is the result of repulsive flux line interactions, where the coherence length is a local property of the superconducting state. Numerous experimental studies have shown \cite{huxley,Brown,Bianchi08}, that the FLL depends sensitively on the microscopic nature, symmetries and number of superconducting gaps, as well as the general topology of the Fermi surface and the effects of impurities. This raises the question how to generalize the description of FLLs and makes studies of conventional superconductors of general interest. 

The superconductivity in Nb is ideally suited to provide such general information \cite{laver_nb,mark_nb2}. The properties of Nb are characteristic of a single gap \cite{Almond,Novotny}, avoiding the complexities of multi-gap systems \cite{laube,huxley,NbSn3}. The low value of the Ginzburg-Landau parameter, $\kappa\approx0.8$, places Nb at the immediate border between type 1 and type 2 superconductivity. This implies that the lower critical field $\mu_0 H_{c1}(T\to0)=120\,{\rm mT}$ is high and the intermediate mixed state (IMS) between the Meissner and the Shubnikov phase is extended, thus making the IMS and the underlying change from attractive to repulsive vortex interactions experimentally accessible  \cite{luk02}. Moreover, the large coherence length of Nb implies that the Fermi surface topology and thus non-local effects are important.

\begin{figure}
\begin{center}
\includegraphics[width=0.42\textwidth]{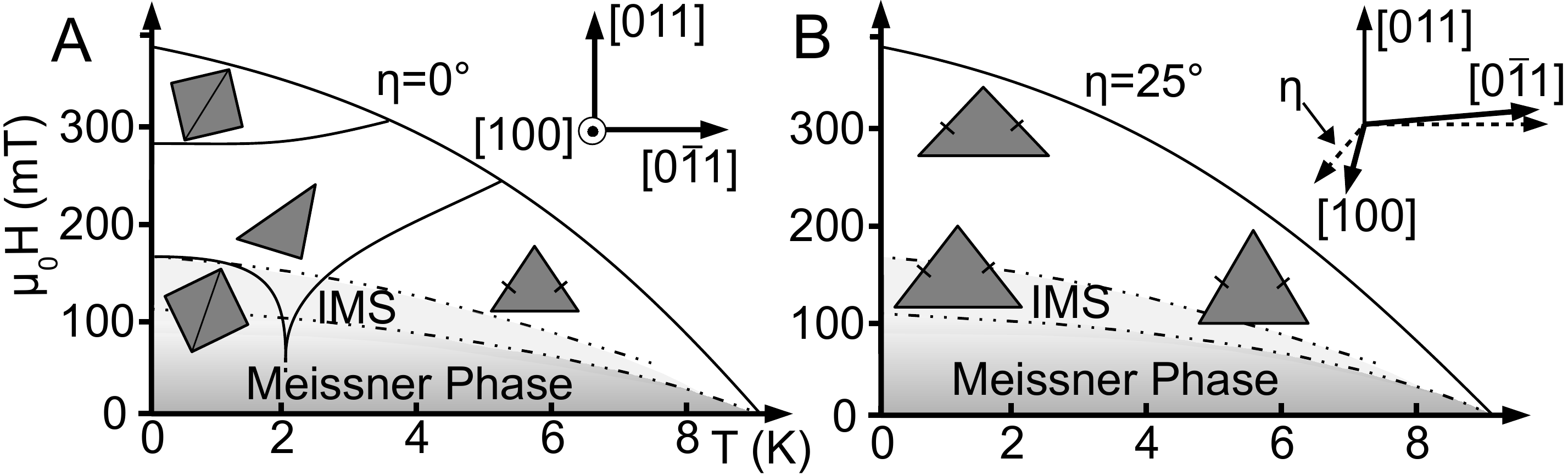}
\caption{(A) Schematic flux lattice phase diagram for a niobium sample with a finite demagnetizing factor in reciprocal space with the magnetic field applied along a four-fold $\left\langle 100 \right\rangle$ axis, where $\eta=0^{\circ}$ denotes the angle between the applied magnetic field and the $\left\langle 100 \right\rangle$ axis in a (011) plane. (B) Phase diagram for $\eta=25^{\circ}$. See text for further details.}
\label{fig1}
\end{center}
\vspace{-8mm}
\end{figure}

For magnetic field $H$ parallel to a crystallographic $\left\langle 111 \right\rangle$ direction the FLL in Nb exhibits a hexagonal symmetry. In contrast, for $H \parallel \left\langle 100 \right\rangle$, a rich phase diagram unfolds shown in Fig.\,\ref{fig1}(A), which prior to our study was understood as follows \cite{laver_nb,Christen80,mark_nb2}: A two-fold isosceles phase and three distinct FLL phases are observed in the vicinity of $T_c$ and at low temperatures, respectively \cite{laver_nb,Christen80,mark_nb2}. For increasing $H$ at low-$T$ these are at first a square phase (low-field square), a scalene phase at intermediate fields and a square phase near $H_{c2}$ (high-field-square). Based on the magnetic field dependence for $\left\langle 100 \right\rangle$ it had been speculated that the low-field-square may be related to the IMS \cite{laver_nb}. It had further been established that the three low-$T$ phases are tilted with respect to the four-fold symmetry of the underlying crystal structure into a low symmetry direction, thus additionally breaking the crystal mirror symmetry. The two-fold isosceles phase in comparison breaks crystal rotational symmetry but still obeys mirror symmetry. It had finally also been established that the FLL symmetries change and the tilting vanishes when rotating the magnetic field away from $\left\langle 100 \right\rangle$ until the equilateral FLL is recovered for $\left\langle 111 \right\rangle$. 

However, in these studies the precise evolution of the FLL morphology as a function of field orientation, necessary to identify the nature of the morphology and tilting, remained open. This originated in incomplete data sets and, more importantly, in the purity of the samples studied so far, which still displayed distinct pinning effects despite their residual resistivity ratios ${\rm RRR}\approx1000$. In this Letter we report a systematic study of the FLL symmetries using small angle neutron scattering (SANS). To resolve the open issues we have studied an ultra-pure single crystal of unprecedented purity with ${\rm RRR}\gtrsim10^4$, which displays essentially no pinning. 

Our study identifies the low-field-square as a property of the IMS. The remaining FLL phases may be viewed as driven by a gradual evolution from an equilateral FLL at high temperature and low field to a square FLL at low temperature and high field (cf Fig.\,\ref{fig1}\,(B)), with a lock-in transition below a magic angle near $\left\langle 100 \right\rangle$, where the locked phases are also tilted. The scalene FLL thereby appears in transitional regions of high frustration. Our experimental results show strong analogies with an evolution of non-local corrections in an Eilenberger \cite{Eilenberger} treatment of a Fermi surface with four-fold symmetry \cite{Nakai} for high-$\kappa$ superconductors. The magic angle of the transition to tilting corresponds remarkably well to a change between open and closed Fermi surface sheets, empirically suspected to affect the superconductivity \cite{bert76}.

For our studies, a cylindrical Nb single crystal with a length of 20\,mm and a diameter of 4.5\,mm was cut from a rod that had been produced at the ZFW Dresden \cite{zfw} more than 30 years ago. The preparation process consisted of purification by liquid-liquid extraction combined with chlorination and thermal decomposition of NbCl$_5$ followed by electron beam floating zone melting, decarburization in oxygen atmosphere and annealing in UHV \cite{bert76,koethe}. The impurity content was estimated to be less than 1\,ppm for interstitial and better than 2\,ppm for substitutional impurities. The sample was again RF-annealed in UHV above $2000^{\circ}$ at the University of Birmingham for one week to remove interstitials followed by surface oxygenation to reduce the Bean-Livingston barrier for surface pinning \cite{laver_nb}. The RRR was measured with an eddy current decay method at the University of Birmingham and found to exceed $\sim10^4$. The AC susceptibility and the magnetization, measured at the Technische Universit\"at M\"unchen were consistent with the literature.

\begin{figure}
\begin{center}
\includegraphics[width=0.31\textwidth]{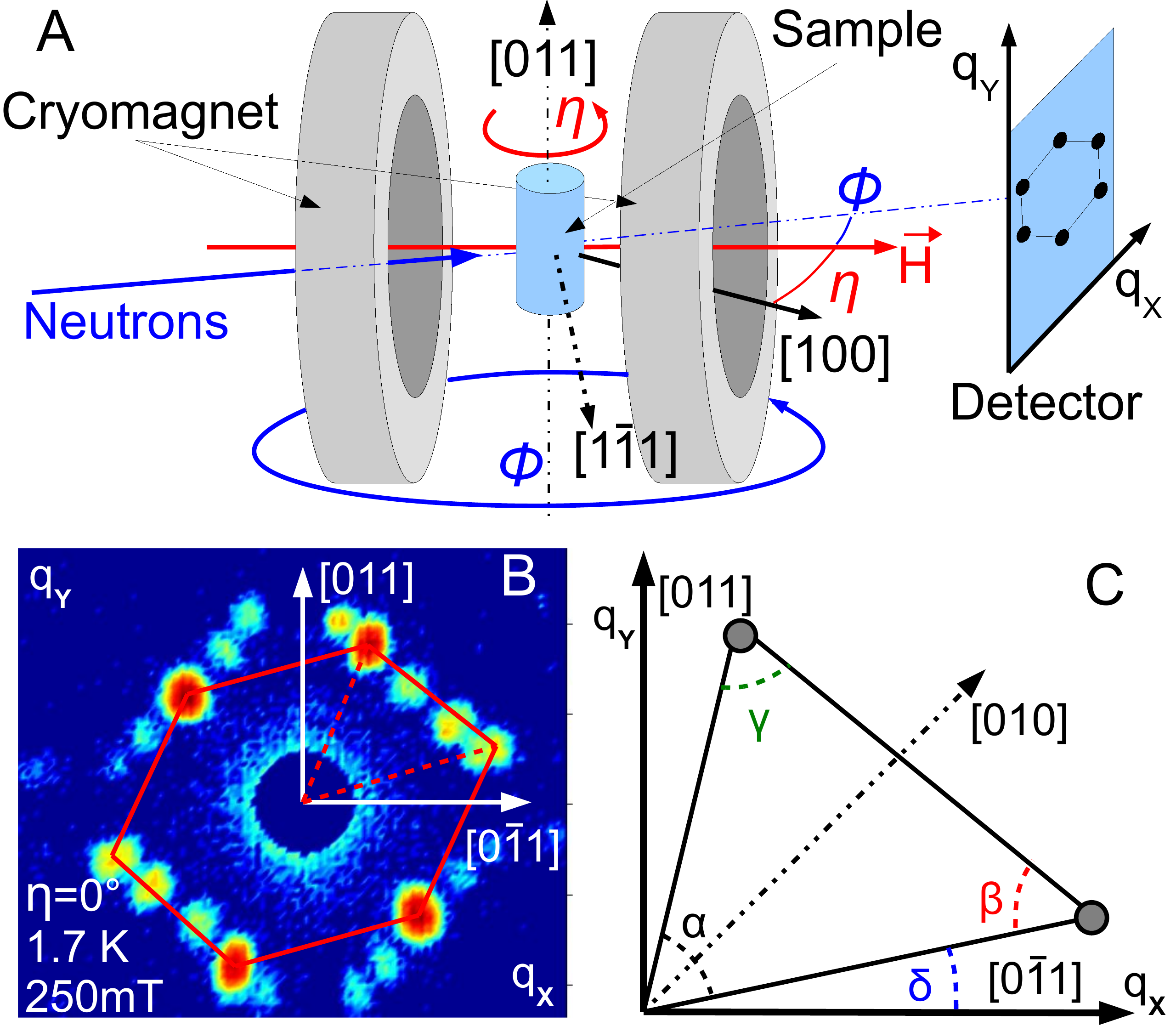}
\caption{(A) Small angle neutron scattering set-up. $\phi$ denotes the angle between the incoming neutron beam and the magnetic field, while $\eta$ denotes the angle between magnetic field and the crystallographic $\langle100\rangle$ direction. (B) Typical detector image in the scalene FLL phase. For clarity just one domain is marked in red. (C) The nomenclature for describing the FLL: $\alpha$, $\beta$ and $\gamma$ represent the internal angles of the FLL unit cell, $\delta$ represents the tilting angle of the FLL with respect to the horizontal $[1\bar{1}1]$ axis. The real space image of the FLL can be obtained by a rotation of the scattering patterns (B) and (C) by 90$^{\circ}$ around the neutron beam axis.}
\label{fig2}
\end{center}
\vspace{-8mm}
\end{figure}

The measurements were carried out at the SANS diffractometer V4 at BENSC. Neutrons with a wavelength $\lambda=12\,{\rm \AA}\pm5.5\%$ were collimated over a distance of 12\,m before reaching the sample, with a multidetector at a distance of 12\,m. An 'orange' cryostat was used for measurements down to 1.5\,K and a horizontal magnet for fields up to 500\,mT. The sample was mounted on a motorized sample stick that could be rotated about the vertical axis with an accuracy of $\pm 0.05^{\circ}$ (cf. Fig.\,\ref{fig2}(A)). The cylinder axis of the sample coincided with the rotation axis and corresponded to a crystalline [011] direction. Magnetic fields were applied in the (011) plane, perpendicular to the cylinder axis so that demagnetizing effects did not change during our measurements. Thus the direction of $H$ could be rotated away from the [100] axis towards a $[1\bar{1}1]$ axis in the (011) plane \cite{rot}. Akin phenomena to those reported here have been obtained for a (100) plane of rotation on a different sample \cite{mark_nb2}. In the following we denote with $\phi$ the rocking-angle between the incoming neutron beam and $H$, while $\eta$ denotes the angle between $H$ and the $\langle100\rangle$ direction.

For the precise sample alignment within $\pm0.2^{\circ}$ we exploited the degeneracy of the superconducting domains for field parallel $\left\langle 100 \right\rangle$ as described in Ref.\,\cite{mark_nb2}. Rocking scans of $\phi$ were performed for each measurement point. To avoid any hysteretic effects, each data point was taken after field cooling the sample \cite{rot}. Several domain populations may be seen for $\eta=0$ at $\mu_0 H=250\,{\rm mT}$ and $T=1.7\,{\rm K}$ (Fig.\,\ref{fig2}(B)), as expected for the symmetry breaking of the FLL with respect to the four-fold symmetry of the crystal lattice. For clarity only one of these domains has been marked with red lines. To extract the morphology of the FLL the scattered intensities were fitted by 2D-Gaussians. As shown in Fig.\,\ref{fig2}(B)(C) $\alpha$, $\beta$ and $\gamma$ describe the internal angles of the scattering pattern, defining the structure of the primitive unit cell, while $\delta$ describes the orientation of the primitive FLL unit cell with respect to the horizontal $[1\bar{1}1]$ axis. Previous studies have shown that changes of domain population of the FLL already occur within small changes of $\eta$, while the shape and orientation of the unit cell are conserved \cite{mark_nb2}. Thus we focus on the morphology of the FLL.

\begin{figure}
\begin{center}
\includegraphics[width=0.34\textwidth]{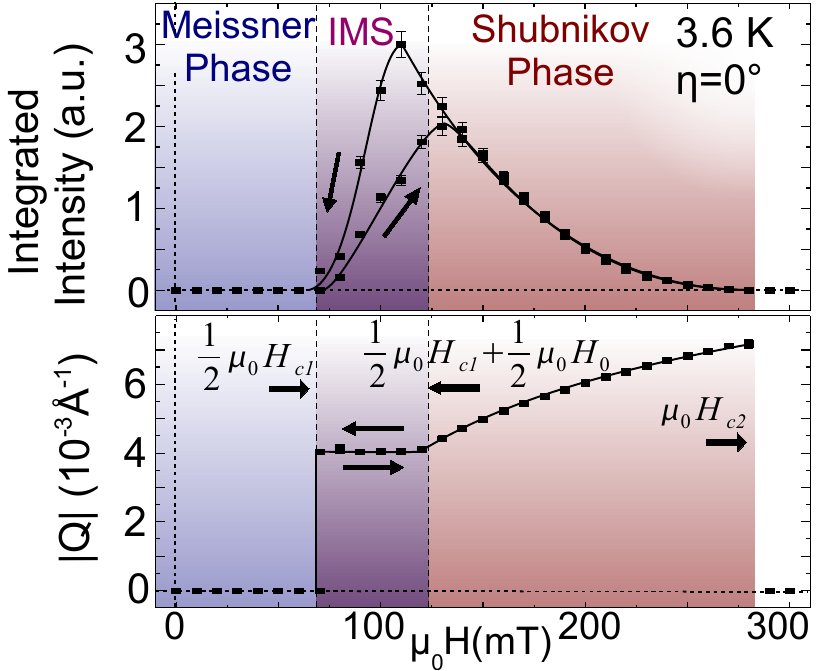}
\caption{Integrated scattering intensity over a rocking curve and reciprocal lattice vector of the SANS diffraction pattern as a function of increasing and decreasing applied magnetic field in the isosceles FLL phase. Only one of two characteristic values of $\left|Q\right|$ is shown, as only rocking curves about a vertical axis could be carried out. Note the complete Meissner flux expulsion for both increasing and decreasing field.}
\label{fig3}
\end{center}
\vspace{-5mm}
\end{figure}

\begin{figure}
\includegraphics[width=0.44\textwidth]{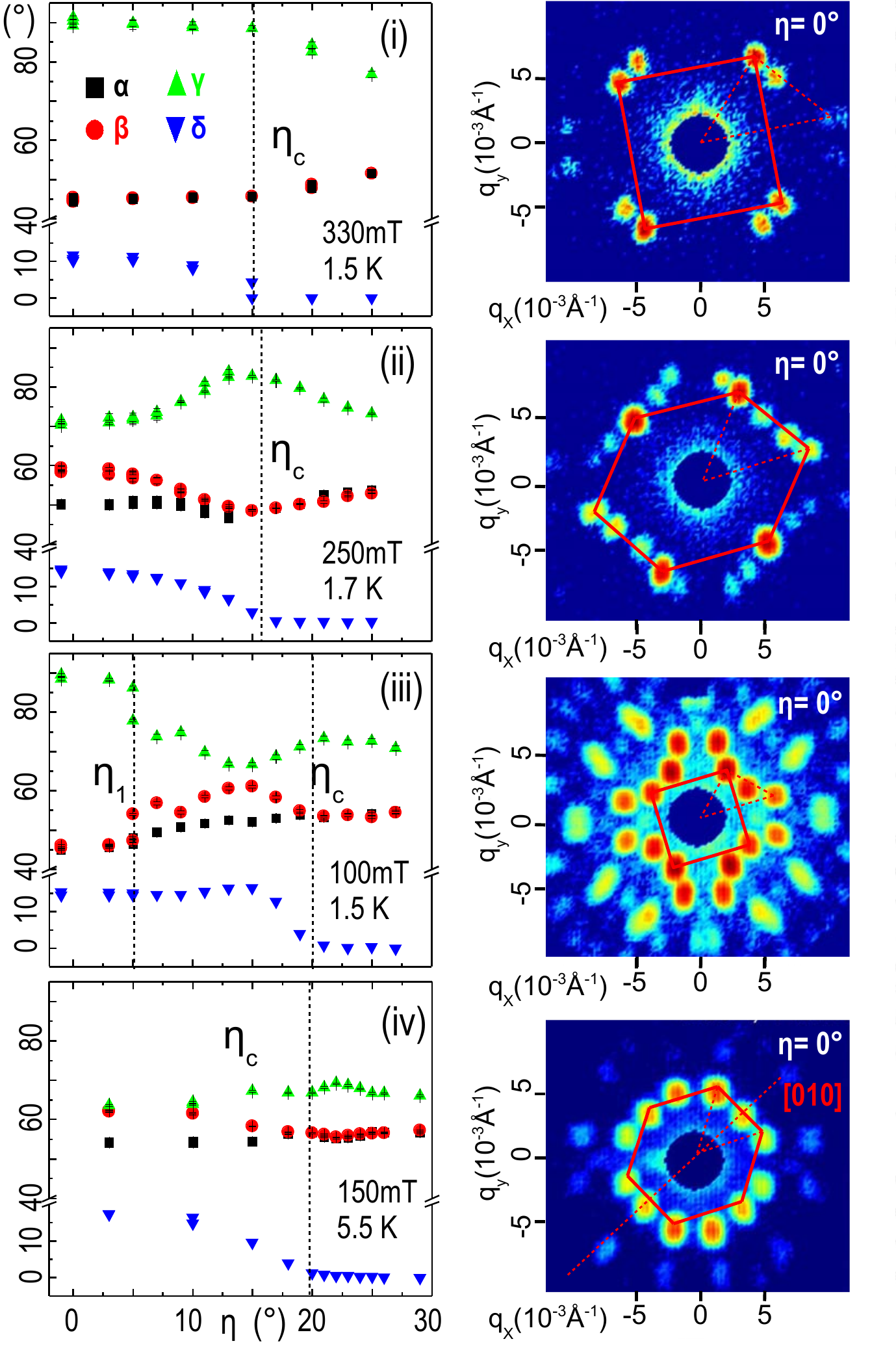}
\caption{Evolution of the flux lattice as a function of $\eta$ for various temperatures and magnetic fields, where typical sums over rocking scans are shown for the high-field-square (i), scalene (ii), low field square (iii) and isosceles (iv) structures. All FLL structures are given in reciprocal space, with the angles plotted in the l.h.s. panels defined in Fig.\ref{fig2}. The scattered intensity is plotted on a logarithmic scale. Due to a slight misalignment of the sample, the scattering patterns are rotated by $\approx$ 1$^{\circ}$ around the beam axis \cite{rot}.}
\label{fig4}
\vspace{-5mm}
\end{figure}

The field dependence of the integrated scattering intensity of a FLL Bragg spot and its reciprocal lattice spacing $\left|Q\right|$ are shown in Fig.\,\ref{fig3} for $T=3.6\,{\rm K}$ and $\eta=0^{\circ}$. We denote the constant flux density due to the attractive flux line interactions in the IMS with $H_0$. As expected for our cylindrical sample with demagnetization factor of $\frac{1}{2}$, no intensity is observed for both increasing and decreasing magnetic field up to $\frac{1}{2}\mu_0 H_{c1}=65\,{\rm mT}$, followed by a gradual increase up to $\frac{1}{2} \mu_0 H_0 + \frac{1}{2}\mu_0 H_{c1}\approx 115\,{\rm mT}$ \cite{Christen}. To the best of our knowledge such a complete Meissner flux expulsion for decreasing $H$ has never been seen before in SANS, underscoring the outstanding purity and perfection of our sample. Above  $\frac{1}{2} \mu_0 H_0 + \frac{1}{2}\mu_0 H_{c1}$ the intensity decreases characteristic of a conventional Shubnikov phase. The value of $Q$ reversibly assumes a constant value $Q\approx0.004\,{\rm \AA^{-1}}$ between $\frac{1}{2}\mu_0 H_{c1}$ and $\frac{1}{2}\mu_0 H_0 + \frac{1}{2}\mu_0 H_{c1}$, identifying this regime as the IMS. In contrast the integrated intensity displays hysteretic behavior between $110\,{\rm mT}$ and $125\,{\rm mT}$. Moreover, also the rocking width of the FLL is increased for decreasing field. The same behavior in sign and magnitude has been observed in the IMS in low quality Nb \cite{Christen}. The data in our high quality sample hence reflect an intrinsic property, most likely due to dendritic flux intrusion for increasing fields in contrast to FLL islands for decreasing fields \cite{Essmann}.

We have studied the evolution of the FLL morphology in detail for $1.5\,{\rm K}<T< 5.5\,{\rm K}$, $110\,{\rm mT}<H< 330\,{\rm mT}$ and $\eta\leq \left|\pm 25^{\circ}\right|$. This covered the parameter range necessary to determine the FLL phase diagram. Shown in Fig.\,\ref{fig4}\,(i) are typical data of the high-field-square for $\eta=0^{\circ}$ with $\alpha=\beta\approx45^{\circ}$, $\gamma\approx90^{\circ}$ and a tilting angle $\delta\approx11^{\circ}$. As a function of rotation angle $\eta$ the symmetry breaking vanishes for $\eta_c\approx15^{\circ}$. Above $\eta_c$ the internal angles $\alpha=\beta$ increase and $\gamma$ decreases characteristic of an isosceles lattice that transforms into the hexagonal lattice with $\alpha=\beta=\gamma=60^{\circ}$ for H $\parallel$ $\langle111\rangle$.

Fig.\,\ref{fig4}\,(ii) shows typical data in the scalene phase, which is characterized by $\alpha\approx 50^{\circ}$, $\beta\approx60^{\circ}$ and $\gamma\approx70^{\circ}$. Like the high-field-square the scalene phase is tilted, where $\delta\approx14^{\circ}$. The tilting vanishes at $\eta_c\approx15^{\circ}$, where the scalene structure turns isosceles, i.e., $\alpha=\beta\approx50^{\circ}$ and $\gamma\approx80^{\circ}$. Approaching $\langle111\rangle$ the FLL turns hexagonal. Typical data of the low-field square are shown in Fig.\,\ref{fig4}\,(iii) ($\alpha=\beta=45^{\circ}$, $\gamma=90^{\circ}$), which abruptly turns into a scalene lattice already for an angle $\eta_1\approx5^{\circ}$, while the tilting of the flux lattice exists up to $\eta_c\approx20^{\circ}$, i.e., in contrast to all other phases the fundamental symmetry and the tilting exhibit distinctly different critical angles.

\begin{figure}
\includegraphics[width=0.46\textwidth]{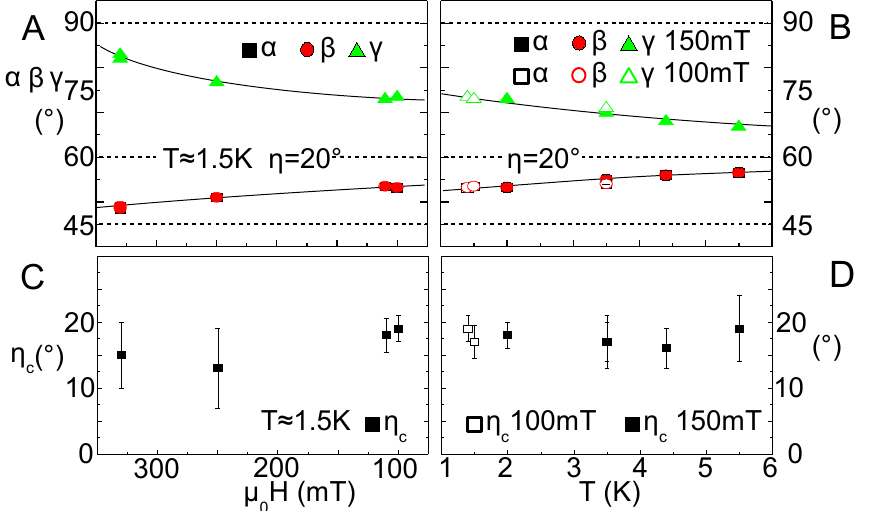}
\caption{Evolution of the internal angles $\alpha$,$\beta$ and $\gamma$ for $\eta=20^{\circ}$ as a function of magnetic field for low temperatures (A) and as a function of temperature for 100\,mT and 150\,mT (B). Note the trend towards a square lattice for low $T$ and high $H$. Panels (C) and (D) show the temperature and field dependence of $\eta_c$, which is essentially unchanged, where the error bars indicate the range of the transition.}
\label{fig5}
\vspace{-6mm}
\end{figure}

Finally, the reorientation angle $\eta_c\approx20^{\circ}$ is also important in the isosceles phase at high temperatures ($\alpha=\beta\approx62^{\circ}$) as shown in Fig.\,\ref{fig4}\,(iv) \cite{Christen80}. For $\eta=0^{\circ}$ the unit cell is orientated such that a [010] axis acts as mirror plane, yielding $\frac{\alpha}{2} +\delta =45^{\circ}$, i.e., $\delta\approx13^{\circ}$. The two-fold isosceles phase hence {\it does} break the underlying four-fold rotational symmetry of the lattice at $\eta=0^{\circ}$ but still obeys mirror symmetry. With increasing $\eta$ this orientation changes between $\eta\approx10^{\circ}$ and $\eta_c$ such that $[01\bar{1}]$ becomes a mirror plane of the isosceles unit cell ($\delta$=0). During this reorientation the unit cell is scalene. This may be compared with the low-field-square shown in Fig.\,\ref{fig4}\,(iii), which becomes scalene already at $\eta_1\approx5^{\circ}$ as discussed above. At $\eta_c\approx20^{\circ}$ both the tilting vanishes and the lattice returns to the isosceles structure.

Thus $\eta_c$ represents a magic angle that is essentially the same for all FLL phases (Fig.\,\ref{fig5} (C) and (D)). This is strongly contrasted by $\eta_1$ of the low-field-square. Moreover, the reversible transition fields of the low-field-square and the IMS shown in Figs.\,\ref{fig3} and \ref{fig4} places the low-field-square right in the IMS. Due to the large $\xi_{GL}$ as compared to the spacing of the flux lines, the interaction between flux lines in the IMS includes contributions from an overlap of the vortices. This leads to a long-range attractive FLL interaction in the IMS, where a square FLL is expected in a theoretical analysis, ignoring any symmetries imposed by the crystal lattice \cite{Kramer}, consistent with our experiment. Taken together this identifies the low-field square as a property of the IMS. 

With the low-field-square being a property of the IMS we may now address the remaining phase diagram. We first consider gradual variations of the FLL morphologies, where it is instructive to begin with the isosceles phase for $\eta>\eta_c$ (cf Fig.\,\ref{fig1}(B)). As shown in Fig.\,\ref{fig5} for $\eta\approx20^{\circ}$ we find that $\alpha=\beta\to45^{\circ}$ and $\gamma\to90^{\circ}$ for decreasing temperature and increasing magnetic field. In other words the lower the temperature and the higher the magnetic field the stronger the trend to form a square lattice. This is underscored by a weak four-fold anisotropy of $H_{c1}$ \cite{Kerchner, Christen80}. The same trend to form a square lattice is also present for $\eta<\eta_c$, where the FLL morphology locks into distinct FLL phases. A scalene FLL thereby emerges always in the transition regime between structures with higher symmetry (isosceles or square), i.e., under conditions of maximum frustration. This may be readily seen in Figs.\,\ref{fig4}\,(ii), (iii) and (iv), where we find that $\alpha$ and $\beta$ differ for $\eta < \eta_c$, while $\alpha = \beta$ for $\eta > \eta_c$. Likewise, a scalene FLL due to frustration is also seen during the rotation between the two isosceles phases in Fig.\,\ref{fig4}\,(iv) with $[01\bar 1]$ and [010] acting as mirror planes, respectively. Hence the same systematic trend is observed for all reorientation transitions shown in Fig.\,\ref{fig4}\,(ii), (iii) and (iv), but where the width of the transition differs.

In a pioneering study Nakai et al \cite{Nakai} have shown for high-$\kappa$ superconductors, that the trend observed here is expected for a four-fold Fermi surface symmetry and related four-fold gap anisotropy. This reproduces qualitatively the behaviour with $H \parallel $ $\langle 100 \rangle$, where a decreasing amount of this trend may be expected to survive when turning towards $\langle 111\rangle$. 

We finally note, that roughly the same value of $\eta_c$ and the size of the tilting, $\delta_c$, are observed across the entire phase diagram. These values are remarkably close to the angle separating open and closed Fermi surface sheets inferred from the magnetoresistance and Fermi surface calculations \cite{Mattheis,Fawcett}, empirically suspected to be important for the superconductivity \cite{bert76}. Moreover, the tilting angle $\delta_c$ compares very well with the size of certain necks in the Fermi surface \cite{Reed}. Thus the trend of the FLL morphology to lock into stable phases wins at the expense of the symmetry breaking rotations as driven by the Fermi surface topology. 

In conclusion, we attribute the general morphology of the FLL in Nb and its orientation to three dominant mechanisms. First, non-local contributions as described by Nakai {\it et al} in the Eilenberger formalism \cite{Nakai}, second, the transition between open and closed Fermi surface sheets and, third, the intermediate mixed state at the border between the Meissner and the Shubnikov phase.

We wish to thank E. H. Brandt, M. D\"aubler, R. Hackl, M. Janoschek, V. Mineev, D. Wallacher, J. White and W. Zwerger for support and stimulating discussions.


\begin{thebibliography}{99}
\bibitem{huxley} A. Huxley et al, Phys. Rev. Lett., {\bf 93}, 187005, (2004).
\bibitem{Bianchi08} A. Bianchi et al, Science {\bf 319}, pp. 177 - 180 11, (2008).
\bibitem{Brown} S. P. Brown et al, Phys. Rev. Lett., {\bf 92}, 067004 (2004).
\bibitem{laver_nb} M. Laver et al, Phys. Rev. Lett., {\bf 96}, 167002, (2006).
\bibitem{mark_nb2} M. Laver et al. Phys. Rev. B, {\bf 79}, (2009).
\bibitem{Almond} D. P. Almond, Phys. Rev. Lett., {\bf 29}, 764, (1972).
\bibitem{Novotny} V. Novotny, J. Low Tem. Phys., {\bf 18}, 147, (1975).
\bibitem{laube} F.Laube et al, Europhys. Lett., {\bf 56} (2), p. 296 (2001). 
\bibitem{NbSn3}E. Boaknin et al, Phys. Rev. Lett. {\bf 90}, 117003 (2003).
\bibitem{luk02} F. Mohamed et al, Phys. Rev. B, {\bf 65}, 224504, (2002).
\bibitem{Christen80} D.K. Christen, Phys. Rev. B, {\bf 21}, 102, (1980).
\bibitem{Eilenberger} G. Eilenberger et al, Z. Phys., {\bf 214}, 195 (1968).
\bibitem{Nakai} N. Nakai et al, Phys. Rev. Lett., {\bf 89}, 237004-1, (2002).
\bibitem{bert76} K.-H- Berthel, dissertation, Dresden (1976).
\bibitem{zfw} ''Zentralinstitut f\"ur Festk\"orperphysik und Werkstoffpr\"ufung'' of the ''Akademie der Wissenschaften'' of the German Democratic Republic, now IFW Dresden.
\bibitem{koethe} A. K\"othe and J. i. M\"onch, Mat. Trans. JIM, {\bf 41}, 7, (2000).
\bibitem{rot} Due to a small sample misalignment of $\approx 1^{\circ}$ around the beam axis, the rotation plane is not exactly a (011) plane.
\bibitem{Christen} D. K. Christen et al, Phys. Rev. B., {\bf 15}, 4506, (1977).
\bibitem{Essmann} U. Essmann, Phys. Lett. A, {\bf 41}, (1972).
\bibitem{Kramer} L. Kramer, Phys. Rev. B, {\bf 3}, 3821, (1971).
\bibitem{Kerchner} H. R. Kerchner et al, Phys. Rev. B., {\bf 24}, 1200, (1981).
\bibitem{Fawcett} E. Fawcett et al, Phys. Rev.,  {\bf 159}, 533, (1967).
\bibitem{Mattheis} L. F. Mattheiss, Phys. Rev. B,  {\bf 1}, 373, (1970).
\bibitem{Reed} W. A. Reed et al, Phys. Rev.,  {\bf 173}, 677, (1968).
\end{thebibliography}
\end{document}